\documentclass{elsart1p}
\usepackage{graphicx}
\usepackage{amssymb}
\begin{document}
\begin{frontmatter}
%
%
%

\title{The minimal two DM components with SUSY\thanksref{label1}}
\thanks[label1]{This work is supported in part by the Korea Research Foundation, Grant No. KRF-2005-084-C00001, and B.K. is also supported by the FPRD of the BK21 program and the KICOS Grant No. K20732000011-07A0700-01110 of the Ministry of Education and Science of Republic of Korea.}

\author{Ji-Haeng Huh, Jihn E. Kim and Bumseok Kyae}
\address{Dept. of Physics and Center for Theoretical Physics, Seoul National University, Seoul 151-747, Korea}
\begin{abstract}
We present the dark matter extension of the minimal supersymmetric standard model by one more stable fermion $N$ toward explaining the recent rising high energy positron spectrum of the PAMELA data. The needed coupling can arise in the flipped-SU(5) GUT. \end{abstract}
\end{frontmatter}

From the early 1930s, dark matter(DM) in the universe has been suggested from the study of the virial mass of galaxy clusters. In the recent years, this has been strengthened by rather reliable measurement on the flat rotation curve of the star velocities in the halo, the simulation of the Bullet Cluster collision, gravitational lensing experiments, etc.

Related to cold dark matter(CDM), a very interesting observation has been reported this summer by the PAMELA satellite experiments, which shows the slightly rising $e^+$ high energy spectrum in the $E\sim 10-60$ GeV region \cite{PAMELAe}, but no such rising hint in the $\bar p$ spectrum \cite{PAMELAp}. In this short talk, we present our works attempting to explain these surprising observations within the low energy SUSY framework \cite{TwoDMs,BaeHuh08}. An easy explanation of the high energy positrons in the particle physics viewpoint is the CDM annihilation producing positrons.

The weakly interacting massive particles(WIMPs) have long been believed to be CDM sources because of their correct order of magnitude for the scattering cross section. Heavy neutrinos and the lightest neutralino $\chi$({LN$\chi$}) have been the major favorites. Most physicists believe in the existence of CDM, but so far we lacked any hint on its nature:
\begin{itemize}
\item What is the mass? 1 MeV?, 100 GeV?, 10 TeV?, or $10^{10}$ GeV?
\item What is the spin? Is it spin-0 axion, sneutrino or darkon?, spin-$\frac12$ neutralino?, spin-$\frac32$ gravitino?, or  spin-1 Kaluza-Klein photon?
\item What are the quantum numbers?
\item Is it stable, or unstable?
\end{itemize}

There have been numerous recent attempts to see the effects of DM in the EGRET, ATIC, PAMELA, HEAT, and INTEGRAL experiments. Before the PAMELA report, there have been reports on the $e^+$ excess in the HEAT, CAPRICE, AMS-01, etc., but those reports were with large error bars and the rising positron flux was not conclusive. The PAMELA report has been remarkable because of the unexpected high energy $e^+$ spectrum with small error bars \cite{PAMELAe}. This rising high energy $e^+$ spectrum may be due to uninteresting astrophysical sources from pulsars or under-estimated systematic errors, but here we focus on the interesting possibility of its particle physics origin.

There are several important issues related to the PAMELA data if it is due to a particle physics candidate $X$ for the CDM:
\begin{itemize}
\item $X$ predominantly decays to $e^+e^-$,
\item $X$ has a small branching ratio to $\bar pp$,
\item For a single component $X$, a large boost factor ($10-10^4$) is needed, and
\item The spin-0 $X$ and the single Majorana DM model (such as the {LN$\chi$} of the minimal supersymmetric standard model(MSSM)) are almost ruled out.
\end{itemize}
Regarding the last item, one may ask, ``Is low energy SUSY dead?" The answer is, ``Not exactly", which we try to explain here.

The PAMELA data, if it is from particle physics source, is not consistent with the one-component {LN$\chi$} with the low energy SUSY and spin-0 models (e.g. darkon). For super-WIMPs(axino, gravitino) with SUSY, they must decay to WIMPs and have the same fate as the {LN$\chi$} of the MSSM. If the super-WIMP is the {LN$\chi$}, it cannot explain the PAMELA high energy $e^+$. The argument is the following. Let $X$ be the Majorana fermion $\chi$. For the direct production of $e^+e^-$ from $\chi\chi$ annihilation, we note for the initial state, (i) the Majorana fermion is self anti-particle, (ii) it has spin-$\frac12$ and obeys the Fermi-Dirac statistics, and (iii) the CDM $\chi$ is very slow today $(v \sim 0.001)$ and hence $\chi\chi$ is in the $s-$wave state. Thus, the initial angular momentum $J=L+S$ is almost zero. Therefore, in the direct production case we note for the final $e^+e^-$, (a) $e^+$ and $e^-$ are much lighter than the CDM mass, and hence the final
$e^+$ and $e^-$ are in the helicity eigenstates, (b) if $e^+$ and $e^-$ are from the same chiral representation, they have the opposite chiralities. Then, the final spin is 1, which is forbidden by the angular momentum conservation rule, and (c) if $e^+$ and $e^-$  are from the opposite chiral representations, a chirality
flipping interaction is needed, i.e. suppressed by the small electron Yukawa coupling \cite{TwoDMs}. As a result, we expect $\langle\sigma v\rangle_{\chi\chi\to e^+e^-}\propto (m_e/m_\chi)^2\sim 10^{-10}$. To circumvent the angular momentum constraint, one may consider producing $\gamma$ in addition to $e^+e^-$ in the final state, but then one needs a much larger boost factor since such cross section will have a $\alpha_{\rm em}/\pi$ factor multiplied.

\begin{figure}[!h]
\resizebox{0.4\columnwidth}{!}
{\includegraphics{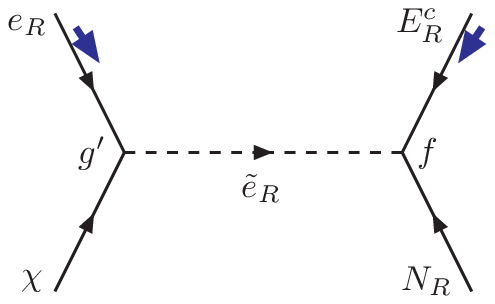}}(a)
\hskip 1.5cm
\resizebox{0.24\columnwidth}{!}
{\includegraphics{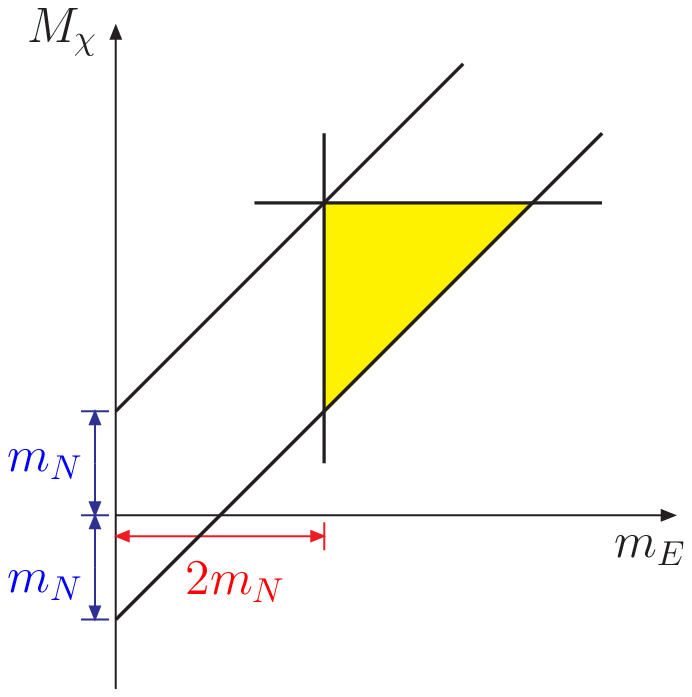}}(b)
\caption{\it  The $\chi-N_R$ annihilation diagram, and the kinematically allowed region in the $M_{\chi}-m_E$ plane. }\label{fig:chiNRscatt}
\end{figure}

\begin{figure}[!h]
\resizebox{0.55\columnwidth}{!}
{\includegraphics{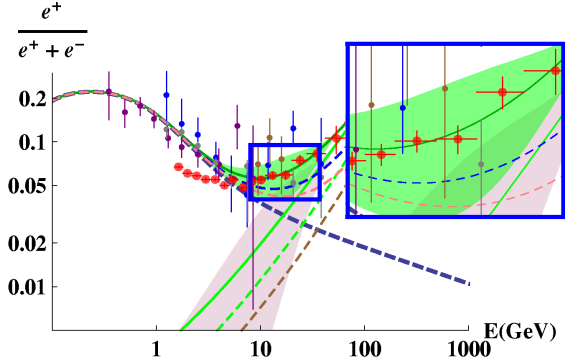}}
\caption{\it The positron fraction from our model with $M_\chi=200$ GeV, $m_N=80$ GeV, $M_{\tilde E}=400$ GeV, $m_E=200$ GeV and $M_{\tilde e}=220$ GeV (thick green line) and $B=7$ \cite{TwoDMs}.}\label{fig:highEpositrons}
\end{figure}

Thus, we consider two DM scenario \cite{TwoDMs}, keeping the low energy SUSY which was the reason to understand the gauge hierarchy problem. The {LN$\chi$} of the MSSM is one CDM component. For the second CDM component, we introduce a neutral chiral superfield $N_R$ with the R-parity +. The $\chi\chi$ annihilation and $N_RN_R$ annihilation have the same angular momentum constraint discussed above, but the annihilation $\chi+N_R\to$ (final state) does not have such a constraint. Because $e_L$ has the weak SU(2) quantum number, the direct annihilation diagram for $\chi+N_R\to e^++e^-$ cannot be drawn. One has to introduce more particles. The minimal extension is to include $Q_{\rm em}=-1$ Dirac fermion. Thus, we add two more particles, $\{N, E\}$, in addition to the MSSM fields, and introduce a superpotential
\begin{equation}
W=fe_R E_R^c N_R.\label{NeEcoup}
\end{equation}
Then, Fig. \ref{fig:chiNRscatt}(a) shows the $e^-$ or $e^+$ production diagram together with $E^+$ or $E^-$. Here $Z_6$ is exact and for the MSSM fields it reduces to the R-parity. With $N$, interestingly, there is a finite region of the kinematic space such that $\chi$ and the neutral fermion $N$ are stable, which is shown in Fig. \ref{fig:chiNRscatt}(b). In Fig. \ref{fig:highEpositrons}, we present the expected high energy positron spectrum with O(100 GeV) CDM candidates \cite{TwoDMs}. Any other extension of the MSSM needs more fields than we suggest here.

The key point of the model is the interaction (\ref{NeEcoup}) where only $N_R$ couples to the singlet $e_R$. Indeed, in the flipped-SU(5) $e_R$ is an SU(5) singlet while $u_R$ and $d_R$ are not SU(5) singlets. This leads to the high energy $e^+$ but no high energy $\bar p$ from the CDM annihilation, consistent with the PAMELA data \cite{PAMELAe,PAMELAp}. It is discussed in more detail in Ref. \cite{BaeHuh08} with the electrophilic axion with the axion decay constant at $F_a\simeq 1.4\times 10^{10}$ GeV.

\end{document}